\documentclass{aa}  
\usepackage{graphicx}
\usepackage{txfonts}
\begin{document}
   \title{Hydrodynamical simulation of detonations in superbursts}

   \subtitle{I. The hydrodynamical algorithm and some preliminary one-dimensional results.}

   \author{C. No\"el
          \inst{1}
          \and
          Y. Busegnies\inst{1}
	  \and
	  M. V. Papalexandris\inst{2}
	  \and
	  V. Deledicque\inst{2}
	  \and
	  A. El Messoudi\inst{1}
          }

   \offprints{C. No\"el}

   \institute{Institut d'Astronomie et d'Astrophysique, Universit\'e Libre de Bruxelles,
              Campus plaine CP 226, Boulevard du Triomphe, 1050 Bruxelles, Belgium\\
              \email{cnoel@ulb.ac.be}
         \and
             D\'epartement de M\'ecanique, Universit\'e catholique de Louvain, 1348 Louvain-la-Neuve, Belgium\\
             }

%   \date{Received September 15, 1996; accepted March 16, 1997}

% \abstract{}{}{}{}{} 
% 5 {} token are mandatory
 
  \abstract
  % context heading (optional)
  % {} leave it empty if necessary  
   {}
  % aims heading (mandatory)
   {This work presents a new hydrodynamical algorithm to study astrophysical detonations. A prime motivation of this development is 
   the description of a carbon detonation in conditions relevant to superbursts, which are thought to result from the  
   propagation of a detonation front around the surface of a neutron star in the carbon layer underlying the 
   atmosphere.}
  % methods heading (mandatory)
   {The algorithm we have developed is a finite-volume method inspired by the original 
  MUSCL scheme of van Leer (1979). The algorithm is of second-order in the smooth part of the flow and avoids 
  dimensional splitting. It is applied to some test cases, and the time-dependent results are compared to the corresponding 
  steady state solution. }
  % results heading (mandatory)
   {Our algorithm proves to be robust to test cases, and is considered to be reliably applicable to astrophysical detonations. 
   The preliminary one-dimensional calculations we have performed demonstrate that the carbon detonation at the surface of a neutron star 
   is a multiscale phenomenon. 
   The length scale of liberation of energy is $10^6$ times smaller than the total reaction length. We show that a multi-resolution approach 
   can be used 
   to solve all the reaction lengths. This result will be very useful in future multi-dimensional simulations. We present also thermodynamical and 
   composition profiles after the passage of a detonation in a pure carbon or mixed carbon-iron layer, in thermodynamical conditions 
   relevant to superbursts in pure helium accretor systems.}
  % conclusions heading (optional), leave it empty if necessary 
   {}


   \maketitle
%
%________________________________________________________________

\section{Introduction}\label{Introduction}

   Superbursts have been discovered by long term monitoring of the X-ray sky with instruments such as RXTE and 
   BeppoSAX. Compared to normal type I X-ray bursts, they are 1000 times more energetic (integrated burst energies 
   of about $10^{42}$ ergs), 1000 times longer (they last from hours to half a day), 
   and have recurrence times of the order of years. They are very rare, only 13 such events having been found from 8 sources
    (for reviews see Kuulkers \cite{Kuulkers2004}, Cumming \cite{Cumming2005}, and references therein). 
   
   They also exhibit similarities with  
   normal type I X-ray bursts, like a rapid rise in the light curve, a quasi-exponential decay, and a hardening of the spectrum during the rise 
   followed by a softening during the decay, which are well represented by a blackbody model with an effective temperature growing 
   during the rise and decreasing during the decay phase (Kuulkers \cite{Kuulkers2004}). This leads to the suggestion that superbursts, 
   like normal type I bursts, are thermonuclear in origin (Cornelisse et al. \cite{Cornelisse2000}). The current view is that 
   superbursts are due to the thermally unstable ignition of $ ^{12}$C at densities of about $10^8$ - 
   $10^9 \, $g cm$^{-3}$ (Cumming \cite{cumming2001}, Strohmayer \& Brown \cite{strohmayer2002}, Cumming \cite{Cumming2005}). 
   
    Most superbursts have been observed in systems accreting a mix of H 
   and He (Kuulkers \cite{Kuulkers2004}).
   In these H/He accretors a very small 
   amount of $ ^{12}$C remains after the combustion of H and He via the rp-process (Wallace \& Woosley 
   \cite{Wallace1981}, Shatz et al. \cite{Schatz2001}). For these systems, Cumming \& Bildsten (\cite{cumming2001}) have shown that a 
   small residual 
   amount of $ ^{12}$C $(X_{ ^{12}\mbox{C}} \approx 0.1)$ in a heavy element bath might be enough to ignite a superburst. 
   Indeed, the low thermal conductivity of the rp-process ashes gives a large temperature gradient in the ocean which favours an  
   unstable ignition of $ ^{12}$C (Cumming \& Bildsten \cite{cumming2001})
   (By ocean, we mean the region far below the zone where the accreted matter is decelerated from its free-fall velocity, and directly underneath 
   the hydrogen/helium burning layer (Brown \& Bildsten \cite{Brown1998})). Moreover Shatz et al. (\cite{Schatz2003})
   have shown that, at the high temperature reached in superbursts ($T>10^9$K), the photodesintegration of the heavy rp-process ashes 
   releases a quantity of nuclear energy which can be larger than the energy release from the fusion of carbon.  
   If this is true, superbursts are the only known cosmic phenomena where photodisintegration of heavy elements is the main energy source.
   This interesting nucleosynthesis aspect will be considered in a future work.
   
   This paper deals only with the case of pure He accretion. 
   Pure He accretors are rare systems. 
   There 
   is only one superburst which has been observed in a system where the 
   accreted material likely has a very high He abundance. This system is 4U 1820-30 (Strohmayer \& Brown \cite{strohmayer2002}, hereafter SB02). 
   It is an ultra compact binary where the companion star 
   is probably a low-mass helium dwarf (SB2002, Cumming \cite{Cumming2003}).
   This superburst detected by RXTE was $\approx$ 3 hours long, had a peak luminosity of $\approx 3.4 \, 10^{38} 
   \, $ergs$ \, $s$^{-1}$, an observed energy release of $1.5 \, 10^{42} \, $ergs, and was preceded by a normal type I burst (20 s duration) 
   (SB02).
   SB02 showed that this superburst is thermonuclear in origin, fueled by the
   burning of carbon produced by the stable burning of the accreted helium between bursts. As neutrinos carry away most of 
   the energy, the total energy of the superburst must be much greater than the observed one ($\approx 10^{44} \,$ergs).
   The ashes of He burning depend on the stability of the combustion. If stable, C is the main product (Brown \& Bildsten \cite{Brown1998}), 
   while an unstable 
   combustion during normal type I bursts 
   produces iron group elements (SB02). SB02 have shown that the iron made during bursts may mix with the carbon made during stable 
   burning between bursts, so that the deep ocean is a mixture of the two. For this reason, we will study here the characteristics 
   of a detonation wave in pure $ ^{12}$C and in a mixture $X_{ ^{12}\mbox{C}} = 0.3$ and $X_{ ^{52}\mbox{Fe}} = 0.7$, which are the limiting 
   cases considered by SB02.
   Due to the lack of protons, no rp-process can develop, which implies the absence of photodisintegration of heavy rp nuclides. In 
   such conditions, a restricted nuclear reaction network is sufficient to describe the nucleosynthesis. This greatly eases the 
   hydrodynamical simulations. 
   We use a 13 species $\alpha$ chain network commonly used in other astrophysical 
   hydrodynamics code (Fryxell et al. \cite{Fryxell1989}).

   Previous works concerning superbursts focused on the thermodynamical state of the surface layers just before ignition, and on the cooling 
   following the bursts (Weinberg et al. \cite{Weinberg2006}), but detailed hydrodynamic calculations are still missing. 
   The observation of oscillations during the 2001 February 22 (UT) superburst from 4U 1636 53 suggests some departure from spherical symmetry 
   in the superburst phenomenon 
   (Strohmayer \& Brown \cite{strohmayer2002b}). Indeed, as it was suggested
   for normal type I X-ray bursts (Shara \cite{Shara1982}), it is unlikely that ignition conditions will be achieved over 
   the entire surface simultaneously. It appears more likely that burning is initiated locally, and then spreads 
   laterally around the neutron star. Since the neutron star is rotating, the oscillations might be understood with a model 
   including rotation and a non-uniform surface brightness, as well as the spreading of the combustion around the surface (Spitkovsky et 
   al. \cite{Spitkovsky2002}). 
   
   A combustion front may propagate in different ways (Williams \cite{williams1965}).
   According to Weinberg et al. (\cite{Weinberg2006}) the superburst rise evolves through three nuclear 
   burning stages: an hour-long convective stage, a runaway stage, and a hydrodynamic stage. A combustion wave forms and 
   may propagate from the site of the runaway as a detonation.  
    
   The only previous numerical studies of the propagation of a detonation front at the surface of a neutron star were made by Fryxell \& 
   Woosley (\cite{Fryxell1982}) and by Zingale et al. (\cite{Zingale2001}). They considered only a detonation propagating in the 
   He layer and not in the underlying C layer of relevance to superbursts. This paper presents the first hydrodynamical simulation of 
   the C-detonation type. This simulation helps illustrate the performance of a new code we have developed. We limit ourselves here 
   to the one-dimensional case. Two-dimensional results will be presented in a forthcoming paper.
   
   Our code is based on a new MUSCL-type parallelized algorithm introduced by Papalexandis et al. (\cite{Papalexandris2002}) and 
   extended to cope with astrophysical conditions. It is described in Sect. \ref{Numericalmethod}. 
   Our time-dependent simulations of detonation in pure $ ^{12}$C or in a mixture of $ ^{12}$C and $ ^{52}$Fe at constant pressure 
   and density typical of superbursts are compared in Sect. \ref{Detonationsprofiles} with steady state predictions. The importance 
   of the resolution is discussed for different initial conditions. Section \ref{Conclusions} 
   contains our conclusions and perspectives.

%__________________________________________________________________

\section{Numerical method}\label{Numericalmethod}

  Our simulations are performed with a modified version of the unsplit, shock-capturing algorithm 
  for multi-dimentional systems of hyperbolic conservations laws with source terms proposed by Papalexandis 
  et al. (\cite{Papalexandris2002}). It is a finite-volume method in the spirit of the original 
  MUSCL scheme of van Leer (\cite{vanLeer1979}). The algorithm is of second-order in the smooth part of the flow. It avoids 
  dimensional splitting. For the purpose 
  of our astrophysical study, the original algorithm is extended to treat a stellar equation of state and a thermonuclear 
  reaction network. We have implemented a Riemann solver based on the one of Colella \& Glaz (\cite{Colella1985}), which  
  is able to treat a general equation of state like an astrophysical one (Fryxell et al. \cite{Fryxell2000}). The parallelization, 
  deemed necessary in order to handle the computational 
  requirements for the problem in hand, is based on the \"{ } mpi \"{ } library, as described in Deledicque \& Papalexandris 
  (\cite{deledicque2005}). 
  
  %__________________________________________________________________
  
  \subsection{Hydrodynamics}\label{Hydrodynamics}
  
  The algorithm solves the adiabatic Euler's equations for compressible, non viscous gas dynamics with source terms in two dimentions. 
  The equations can be 
  written in conservative form as

  \begin{equation} \label{system}
  \frac{\partial \bf{U}}{\partial t}+{\bf{\nabla}}\cdot \bf{F}(\bf{U})=\bf{G}(\bf{U}) \mbox{,}
  \end{equation}
  
  \noindent where
  
  \begin{equation} \label{system2}
  {\bf{U}} = \left( 
  \begin{array}{c}
   \rho\\ 
   \rho {\bf u} \\ 
   \rho e_t \\
   \rho Y_i
   \end{array} 
   \right)
   \,\, , \,\,
   {\bf{F}(\bf{U})} = \left( 
  \begin{array}{c}
   \rho {\bf u}\\ 
   \rho {\bf u}^2 +p \\ 
   {\bf u}(\rho e_t+p) \\
   \rho Y_i {\bf u}
   \end{array} 
   \right)
   \,\, \mbox{and} \,\,
   {\bf{G}(\bf{U})} = \left( 
  \begin{array}{c}
   0\\ 
   0 \\ 
   \rho \varepsilon^{nuc} \\
   \rho R^{nuc}_i  
   \end{array} 
   \right)
   \mbox{.}
  \end{equation}

  \noindent To close the system, we need an equation of state of the form  

	\begin{eqnarray}\label{eos}
	p & = & p (\rho,T,{\bf {Y}}) \mbox{,} \label{eos21} \\
	e & = & e (\rho,T,{\bf {Y}}) \mbox{.} \label{eos22}
	\end{eqnarray}

  \noindent In the equations above, $\rho$ is the density, ${\bf u}=(u,v)$ is the velocity vector, $p$ is the pressure, % $c$ is the adiabatic sound speed, 
  $e_t = e + \frac{{\bf u}^2}{2}$ is the sum of the specific internal energy and the specific kinetic energy, 
  $Y_i$ is the molar fraction of species $i$ \footnote{Note that the molar fraction of the 
  leptonic species $Y_L$ does not appear in equations \ref{eos21}-\ref{eos22}. In the thermodynamic conditions relevant to superbursts, the plasma 
  can be considered as totally ionized. The condition of electroneutrality allows the calculation of $Y_L$ once the $Y_i$\'{}s are known.
  (Cox \& Giuli \cite{Cox1968})},
   $\bf {Y}$ is the vector of $Y_i$, $T$ is the temperature and $\varepsilon^{nuc}$ is 
  the total rate of thermonuclear energy released per gram of matter. 
  For a general reaction as $c_iI+c_jJ \rightleftharpoons c_kK+c_lL$, where $c_{i,j,k,l}$ are the stoechiometric coefficients of 
  the nuclides $I$, $J$ ,$K$ and $L$, $\varepsilon^{nuc}$ 
  and $R^{nuc}_i$ are defined by

	\begin{equation}
	\varepsilon^{nuc} = -N_A \sum_{i=0}^{n_{sp}}M_ic^2  R^{nuc}_i \mbox{,}
	\end{equation}
	and
	\begin{eqnarray}
	R^{nuc}_i & = & \sum_k a_i(k)\lambda_k Y_k + \sum_{j,k,l}\{a_i(i,j)[i,j]_k Y_i^{c_i} Y_j^{c_j} + \nonumber \\
	& & a_i(k,l)[k,l]_i Y_k^{c_k} Y_l^{c_l}\} +
	a_i(k,k,k)\rho^2 N_A^2\langle\sigma v\rangle_{3k}Y_k^3 \mbox{,}
	\end{eqnarray}
	
  \noindent where$[i,j]_k = \rho N_A \langle\sigma v\rangle_{i+j\rightarrow k} \mbox{,}$
  $N_A$ is the Avogadro number, $n_{sp}$ is the number of species in the system and $M_ic^2$ is the rest mass energy of species $i$.
  The notation $\langle\sigma v\rangle_{i+j\rightarrow k}$ represents the thermonuclear reaction rate of the process $i+j \rightarrow k+l$ per pair of 
  particles ($i$,$j$). The quantities $a_i$ are statistical factors determined by the $c_{i,j,k,l}$ (Arnett \cite{arnett1996}).
  
  \subsection{The algorithm}\label{algorithm}
  
  \noindent Equations (\ref{system}) and (\ref{system2}) take the integral form

	\begin{equation} \label{densint}
	\frac{d}{d t}\int_V\rho dV + \int_S \rho {\bf u} \cdot d{\bf S} =0 \mbox{,}
	\end{equation}
	\begin{equation} \label{vitint}
	\frac{d}{d t}\int_V \rho {\bf u} dV + \int_S \rho {\bf u}{\bf u} \cdot d{\bf S} +\int_S p d{\bf S}=0 \mbox{,}
	\end{equation}
	\begin{equation} \label{enerint}
	\frac{d}{d t}\int_V \rho e_t dV  + \int_S \rho e_t{\bf u} \cdot d{\bf S} +\int_S p {\bf u}\cdot  
	d{\bf S} - \int_V \rho \varepsilon^{nuc} dV=0 \mbox{,}
	\end{equation}
	\begin{equation} \label{Yint}
	\frac{d}{d t}\int_V \rho Y_i dV  + \int_S \rho Y_i{\bf u} \cdot d{\bf S} - \int_V \rho R^{nuc}_i dV=0 \mbox{.}
	\end{equation}

  \noindent These equations are written for an arbitrary control volume $V$ whose boundary $S$ has zero velocity. 
  The hydrodynamical part of these equations is treated in the same way as in Papalexandris et al. (\cite{Papalexandris2002}).
  The nuclear part of the system requires a specific treatment, however. 
  The original algorithm of Papalexandris et al. (\cite{Papalexandris2002}) is indeed
  able to treat a single stiff source term. Instead our nuclear reaction network which comprises 27 reactions (see Sect. 
  \ref{eos_network}) introduces a set of stiff differential equations, 
  and so very different time scales. This requires the adoption of a time-splitting version 
  of the algorithm. We keep avoiding the dimensional splitting. 
  The time splitting is of the Strang type (Strang \cite{strang1968}). 
  In this case the system of equations (\ref{system}-\ref{system2}) is solved by the split scheme

  	\begin{equation}\label{strang}
	{\bf{U}}^{n+1}=\mathcal{L}_s^{\Delta t/2}\mathcal{L}_f^{\Delta t}\mathcal{L}_s^{\Delta t/2}({\bf{U}}^n) \mbox{,}
	\end{equation}
  
   \noindent where $\bf{U}^{n+1}$ is the solution at time $t+\Delta t$. 
   Here $\mathcal{L}_f$ is the numerical solution operator for the corresponding homogeneous conservation law 
  
  	\begin{equation}\label{Lf}
	\frac{\partial}{\partial t}\bf{U}+{\bf{\nabla}}\cdot\bf{F}(\bf{U})=0 \mbox{,}
	\end{equation}
  
  \noindent and $\mathcal{L}_s$ is the numerical solution operator for the system of ordinary differential equations 
  
  	\begin{equation}\label{Ls}
	\frac{d}{d t}\bf{U}=\bf{G}(\bf{U}) \mbox{.}
	\end{equation}

  \noindent It is obtained from the semi-implicit extrapolation method of Bader \& Deuflhard (\cite{Bader1983}) which is 
  used to solve the nuclear part of the system of equations (\ref{densint}-\ref{Yint}).

  The procedure 
  of discretization and numerical evaluation of the hydrodynamical part of the integrals (\ref{densint}-\ref{Yint}) at each 
  computational cell is the same as in 
  Papalexandris et al. (\cite{Papalexandris2002}). However the gamma-law Riemann solver of the initial algorithm is replaced 
  by one based on the method of Colella 
  \& Glaz (\cite{Colella1985}) able to treat a general equation of state of the form given by equations (\ref{eos21}-\ref{eos22}).
	
  The numerical scheme, which evaluates the solution at time $(n+1)\Delta t$ from the solution at the previous 
  time $n\Delta t$ for the hydrodynamical part of the system (eq. \ref{Lf}) can be written as

	\begin{eqnarray}
	(m_{i,j})^{n+1} & = & (m_{i,j})^{n} - \Delta t[(l{\bf n}_S \cdot {\bf F}_m)_{i+1/2,j}^{n+1/2}-  
	(l{\bf n}_S \cdot {\bf F}_m)_{i-1/2,j}^{n+1/2}] -  \nonumber \\
	& & \Delta t 
	[(l{\bf n}_S \cdot {\bf F}_m)_{i,j+1/2}^{n+1/2}-(l{\bf n}_S \cdot {\bf F}_m)_{i,j-1/2}^{n+1/2}] \mbox{,}
	\end{eqnarray}
	\begin{eqnarray}
	(m_{i,j} u_{i,j})^{n+1} & = &  (m_{i,j}u_{i,j})^{n} - \nonumber \\
	& & \Delta t[(l{\bf n}_S \cdot {\bf F}_u)_{i+1/2,j}^{n+1/2}-
	(l{\bf n}_S \cdot {\bf F}_u)_{i-1/2,j}^{n+1/2}] - \nonumber \\
	 & & \Delta t 
	[(l{\bf n}_S \cdot {\bf F}_u)_{i,j+1/2}^{n+1/2}-(l{\bf n}_S \cdot {\bf F}_u)_{i,j-1/2}^{n+1/2}] \mbox{,}
	\end{eqnarray}
	\begin{eqnarray}
	(m_{i,j} v_{i,j})^{n+1} & = &  (m_{i,j}v_{i,j})^{n} - \nonumber \\
	& & \Delta t[(l{\bf n}_S \cdot {\bf F}_v)_{i+1/2,j}^{n+1/2}-
	(l{\bf n}_S \cdot {\bf F}_v)_{i-1/2,j}^{n+1/2}] - \nonumber \\
	& & \Delta t 
	[(l{\bf n}_S \cdot {\bf F}_v)_{i,j+1/2}^{n+1/2}-(l{\bf n}_S \cdot {\bf F}_v)_{i,j-1/2}^{n+1/2}] \mbox{,}
	\end{eqnarray}
	\begin{eqnarray}
	(m_{i,j} e_{t \, i,j})^{n+1} & = & (m_{i,j}e_{t \, i,j})^{n} - \nonumber \\
	& & \Delta t[(l{\bf n}_S \cdot {\bf F}_e)_{i+1/2,j}^{n+1/2}-
	(l{\bf n}_S \cdot {\bf F}_e)_{i-1/2,j}^{n+1/2}] - \nonumber \\
	& & \Delta t 
	[(l{\bf n}_S \cdot {\bf F}_e)_{i,j+1/2}^{n+1/2}-(l{\bf n}_S \cdot {\bf F}_e)_{i,j-1/2}^{n+1/2}] \mbox{,} 
	\end{eqnarray}
	\begin{eqnarray}
	(m_{i,j} Y_{i,j,k})^{n+1} & = & (m_{i,j}Y_{i,j,k})^{n} - \nonumber \\
	& & \Delta t[(l{\bf n}_S \cdot {\bf F}_{Y_k})_{i+1/2,j}^{n+1/2}-
	(l{\bf n}_S \cdot {\bf F}_{Y_k})_{i-1/2,j}^{n+1/2}] - \nonumber \\
	& & \Delta t 
	[(l{\bf n}_S \cdot {\bf F}_{Y_k})_{i,j+1/2}^{n+1/2}-(l{\bf n}_S \cdot {\bf F}_{Y_k})_{i,j-1/2}^{n+1/2}] \mbox{,} 
	\end{eqnarray}
	
  \noindent where $l$ and ${\bf n}_S$ are the length of a cell interface and the unit vector normal to a cell interface, respectively.
  The flux vectors are given by
  
	\begin{equation}
	{\bf F}_m \equiv [\rho u,\rho v] \mbox{,}
	\end{equation}
	\begin{equation}
	{\bf F}_u \equiv [\rho u^2 +p,\rho u v] \mbox{,}
	\end{equation}
	\begin{equation}
	{\bf F}_v \equiv [\rho u v,\rho v^2 +p] \mbox{,}
	\end{equation}
	\begin{equation}
	{\bf F}_e \equiv [\rho e_t u + p u,\rho e_t v + p v] \mbox{,}
	\end{equation}
	\begin{equation}
	{\bf F}_{Y_i} \equiv [\rho Y_i u,\rho Y_i v] \mbox{.}
	\end{equation}

  All the numerical simulations presented
  in this paper are performed with a Courant number (Leveque \cite{Leveque1999}) CFL = 0.3, 
  and the temperature is not allowed to change by more than 10\% during a time step (Fryxell et al. \cite{Fryxell1989}).
  
  %__________________________________________________________________
  \subsection{Equation of state and nuclear reaction network} \label{eos_network}

  In the thermodynamical conditions relevant to the surface of a neutron star, the plasma can be considered as fully ionized. 
  Our equation of state accounts for partially degenerate and partially relativistic electrons and positrons. 
  The ions are treated as a Maxwell-Boltzmann gas, and the radiation, considered to be at local thermodynamic equilibrium 
  with the matter, follows the Planck law. 
  We use a tabulated equation of state in the spirit of Timmes\'{}s one (Timmes \& Arnett \cite{Timmesarnett1999}). 
  Coulomb interactions of the bare nuclei with the surrounding electron-positron gas are not taken into account here, 
  and will be included in a future work.\\
  
  The selected nuclear reaction network is the one usually adopted in astrophysical hydrodynamics simulations
  in order to provide an energy source representative of explosive helium and carbon burning in absence of hydrogen. It involves 13 
  nuclides ($ ^4$He,$ \, ^{12}$C,$ \, ^{16}$O,$ \, ^{20}$Ne,$ \, ^{24}$Mg,$ \, ^{28}$Si,$ \, ^{32}$S,
  $ \, ^{36}$Ar,$ \, ^{40}$Ca,$ \, ^{44}$Ti,$  \,^{48}$Cr,$  \,
  ^{52}$Fe and $ ^{56}$Ni) linked by 27 reactions comprising the 11 $(\alpha,\gamma)$ reactions from
  $ ^{12}$C$(\alpha,\gamma)$ $^{16}$O to $ ^{52}$Fe$(\alpha,\gamma)$ $^{56}$Ni, the corresponding 11 endothermic 
  photodesintegrations, the three heavy-ion reactions $ ^{12}$C($ ^{12}$C,$\alpha$)$ ^{20}$Ne, 
  $ ^{12}$C($ ^{16}$O,$\alpha$)$ ^{24}$Mg and $ ^{16}$O($ ^{16}$O,$\alpha$)$ ^{28}$Si, and the triple-alpha reaction 
  and its inverse. As in Fryxell et al. (\cite{Fryxell1989}), the reaction rates are taken from  
  Thielemann et al. (\cite{Thielemann1986}), where each reaction rate is given in the temperature interval 
  $10^8 \leq T \leq 10^{10}$K by
  
	\begin{eqnarray}
	 N_A \langle\sigma v\rangle & = & exp(c_1+c_2 T_9^{-1}+c_3T_9^{-1/3}+c_4T_9^{1/3}+ \nonumber \\
	 & & c_5T_9+c_6T_9^{5/3}+c_7ln(T_9)) \mbox{,}
	\end{eqnarray}
	
  \noindent where $T_9=10^9$K and the values of the numerical coeficients $c_k$ are given by the authors. The network 
  equations are constructed as described by Eq. (\ref{Ls}), and are solved with the use of the variable 
  order Bader-Deuflhard semi-implicit time integrator (Bader \& Deuflhard \cite{Bader1983}) suggested by Timmes (\cite{Timmes1999}). The 
  way this method is introduced in the algorithm is the same as in Press et al. (\cite{numrec}).
  
  %__________________________________________________________________
  \subsection{Validation tests}
  The algorithm  has already been validated for terrestrial detonations with a gamma law equation of state, and a single chemical reaction 
  kinetic represented by an simple Arrhenius law 
   (Papalexandis et al. \cite{Papalexandris2002}).
  Its validity in astrophysical situations with a general equation of state and a nuclear reaction network has been checked with 
  some validation tests considered by Fryxell et al. (\cite{Fryxell1989}). They involve non reactive 
  and reactive shock tubes with the astrophysical equation of state and the nuclear reaction network.  
  Our results are similar to those of Fryxell et al. (\cite{Fryxell1989}). 
  
  The comparaison of our time-dependent results with a steady state calculation, as presented in Sect \ref{Detonationsprofiles}, 
  also validate the accuracy of our algorithm.

  %__________________________________________________________________
  \section{Detonation profiles}\label{Detonationsprofiles}

  One-dimensional steady-state calculations (i.e. calculations where the detonation speed remains constant) provide the main 
  parameters characterizing a detonation, such as  
  the characteristic time and length-scales and the reaction-zone structure. These basic detonation properties are necessary 
  to set the initial parameters and boundary conditions in the time-dependent calculations. 
  The treatment of the steady-state case is called the ZND model (Fickett \& Davis \cite{Fickett1979}). 
  According to this model, the detonation consists of an infinitely thin shock followed by a burning zone. All the reactions 
  take place inside this zone and the released energy sustains the shock. In the laboratoy frame, these steady-state equations are

  \begin{equation}\label{ZND1}
  \frac{d \rho}{dt}=\frac{\Phi}{(D-u)^2-a_f^2} \mbox{,}
  \end{equation}
  \begin{equation}\label{ZND2}
  \frac{d e}{dt}=\frac{P}{\rho^2}\frac{d \rho}{dt}+\varepsilon^{nuc} \mbox{,}
  \end{equation}
  \begin{equation}\label{ZND3}
  \frac{dZ}{dt}=D-u=\frac{\rho_0}{\rho} \mbox{,}
  \end{equation}
  and
  \begin{equation}\label{ZND4}
  \frac{dY_i}{dt}=R^{nuc}_i \mbox{,}
  \end{equation}
  
  \noindent where $D$, $Z$ and $\rho_0$ are the detonation speed, the distance behind the shock and the density of the unburnt matter.
  The quantity $\Phi = \varepsilon^{nuc}(\partial p/\partial e)_{\rho,Y_i}$ is the thermicity, $a_f=(\partial p/ \partial \rho)_{S,Y_i}^{1/2}$ 
  is the frozen sound speed and $S$ is the entropy  (Khoklov \cite{Khoklov1989}). The system of equations (\ref{ZND1}-\ref{ZND4})
   is closed by the equation of state (eqs. \ref{eos21}-\ref{eos22}).
  For a given detonation speed, we compute a post-shock state using the Hugoniot relations (Khoklov \cite{Khoklov1988}) and we 
  integrate from the shock to the end of the burning zone (when the $R^{nuc}_i$ vanish) using equations (\ref{ZND1}-\ref{ZND4}).
  
\begin{figure}
	\resizebox{\hsize}{!}{\includegraphics{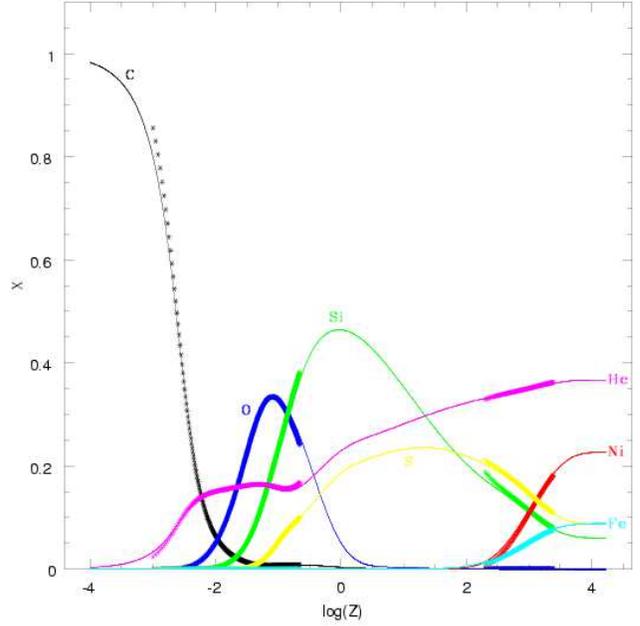}} 
	\caption{ Nuclear mass fraction profiles of $ ^4$He,$ \, ^{12}$C,$ \, ^{16}$O,$ \, ^{28}$Si,$ \, ^{32}$S,$  \,^{52}$Fe and $ ^{56}$Ni 
	for a detonation front in pure $ ^{12}$C at 
	$T=10^8 \, $K and $\rho= 10^8 \, $g$ \, $cm$^{-3}$. $Z$ is the distance to the shock in cm. The thin solid lines give the 
	steady state solution of the ZND model. The heavy dots (the density of the dots is so high that they form thick curves) are obtained with the time-dependent calculations with two 
	different resolutions (mesh size $10^{-4}$ cm and 10 cm).}
	\label{ZND_godunov}
	\end{figure}

  Since superbursts in pure He accreting systems may be due to pure C detonation at high density and temperature we calculate the ZND 
  profiles for a detonation in pure $ ^{12}$C at a temperature $T=10^8 \,  $K and a density 
  $\rho=10^8  \, $g$ \, $cm$^{-3}$. The nuclear mass fraction profiles of some of the most abundant species are presented in 
  Fig. \ref{ZND_godunov} (thin solid lines).
  As can be seen 
  a large variety of length scales are at work. The total reaction length given by the ZND model is of the order of $10^4$ cm.
  Significant changes in the nuclear mass fractions occur already at 
  $10^{-4}$cm. This suggests that the full resolution of the detonation requires a time-dependent simulation over a 
  domain as large as $10^4$ cm with a resolution of $10^{-4}$ cm. Owe to computational time 
  limitations we are unable to reach this resolution. However, following Gamezo et al. (\cite{gamezo1999}), we have been able to perform
   two sets of calculations, one with a resolution of 
  10 cm over a domain of $10^4$ cm, and one with a resolution of $10^{-4}$ cm over a domain of 1 cm.
  
  The system of length $l$ = $10^4 \, $cm is considered
  with an inflow boundary condition at $x = 0$ cm and an outflow boundary condition at $x=$$10^4 \, $cm. As initial 
  conditions, we select pure $ ^{12}$C 
  at $T=10^8 \,$  K, 
  $\rho=10^8  \,$ g cm$^{-3}$ and a material velocity of $0  \,$ cm  $\,$s$^{-1}$. To trigger the detonation we set the initial 
  conditions in an ignition zone between $x = 0$ cm and $x = $$10^3 \, $cm to a temperature of $4.46  \, 10^{9} \, $K, a density 
  of $3.01  \, 10^8 \,  $g$  \, $cm$^{-3}$, a material velocity of $8.07  \, 10^8 \, $cm$  \, $s$^{-1}$ and pure $ ^{56}$Ni.
  We take 1000 numerical cells in the domain, 
  leading to a resolution of 1cm.

  The simulation 
  of a detonation of pure $ ^{12}$C in the same thermodynamic conditions as above 
  has also been performed on a 
  much smaller domain of lenght l = 1cm with $10^4$ numerical cells, so that a resolution of $10^{-4} \, $cm is achieved. 
  The initial discontinuity is positioned at $0.1$ cm.
  
  We have superimposed the profiles of the nuclear mass fractions of some of the most abundant species as a function of the  
  distance to the shock obtained with the ZND algorithm and with the hydrodynamic algorithm. The ZND profiles are obtained for 
  a velocity of propagation of the detonation wave $D=1.3 \, 10^9 \,$ cm s$^{-1}$, which is the velocity of the detonation front in the 
  time-dependent hydrodynamic simulation. As can be seen in Fig. \ref{ZND_godunov}, the time-dependent profiles are very close to the steady state ones. 
  So, as in Gamezo et al. (\cite{gamezo1999}), the partial resolution approach can be applied for the simulation of detonations in 
  this system. This will be very useful in future 2D calculations where the computational time is crucial.

\begin{figure}
	\resizebox{\hsize}{!}{\includegraphics{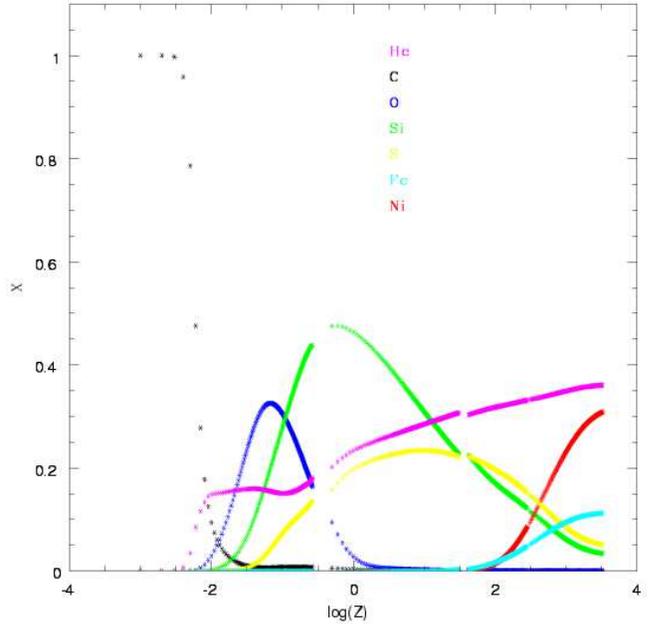}} 
	\caption{Same as Fig. \ref{ZND_godunov}, but with screening effects taken into account.  
	Results only obtained with the time-dependent calculations 
	for four 
	resolutions (from left to right: mesh size $10^{-3}$ cm, 0.1 cm, 1 cm and 10 cm).}
	\label{nucl_C_screen}
   \end{figure}

  The preceding comparison between the steady-state and the time-dependent results 
  was made without taking nuclear reaction rate screening effects into account. 
  However, at high densities, screening may be important (Cox \& Giuli \cite{Cox1968}).
  and all the following results are obtained 
  with the adoption of the screening corrections of Wallace et al. (\cite{Wallace1982}). With this correction we have performed 
  the simulation 
  of a detonation of pure $ ^{12}$C in the same thermodynamic conditions as above at four different resolutions: one 
  with a domain of length l = 1cm, one with l = 100cm, 
  one with l = 1000cm, and one with l = $10^4$ cm, always using 
  $1000$ numerical cells. The four corresponding nuclear mass fraction 
  profiles are presented in Fig. \ref{nucl_C_screen}. 
  One sees the impact of the screening effects by comparing 
  Figs. \ref{ZND_godunov} and \ref{nucl_C_screen}. Screening decreases the reaction length, increases the detonation 
  velocity, and modifies the final composition. However these effects are small (less than 1\% of the detonation 
  velocity).

\begin{figure}
	\resizebox{\hsize}{!}{\includegraphics{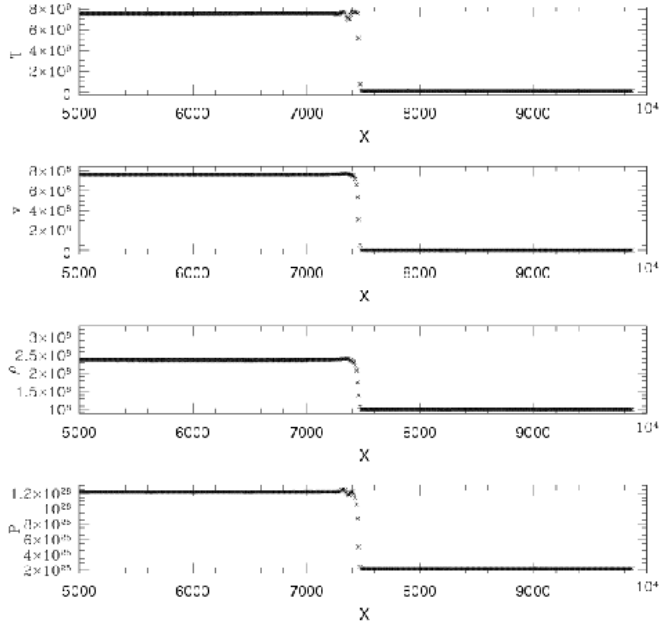}} 
	\caption{Temperature (in K), velocity (in cm$ \, $s$^{-1}$), density (in g$ \, $cm$^{-3}$) 
	and pressure (in erg$ \, $cm$^{-3}$) profiles of a detonation front in pure $ ^{12}$C 
	at  $T=10^8 \, $K and $\rho= 10^8 \, $g$ \, $cm$^{-3}$ at time = $5 \, 10^{-6}$s. $X$ is in cm.}
	\label{therm_C_1e4_screen}
	\end{figure}

\begin{figure}
	\resizebox{\hsize}{!}{\includegraphics{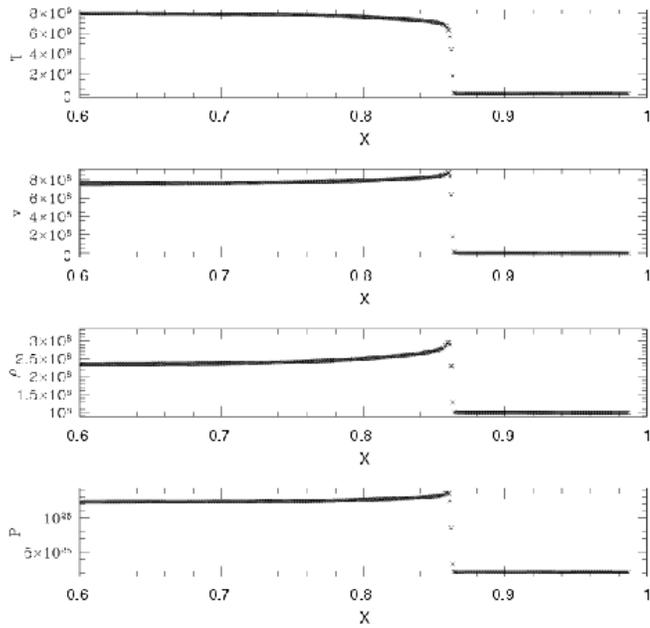}}
	\caption{Same as Fig. \ref{therm_C_1e4_screen}, but at time = $6 \, 10^{-10}$s with a resolution of $10^{-3}$cm.}
	\label{therm_C_1_screen}
	\end{figure}

\begin{figure}
	\resizebox{\hsize}{!}{\includegraphics{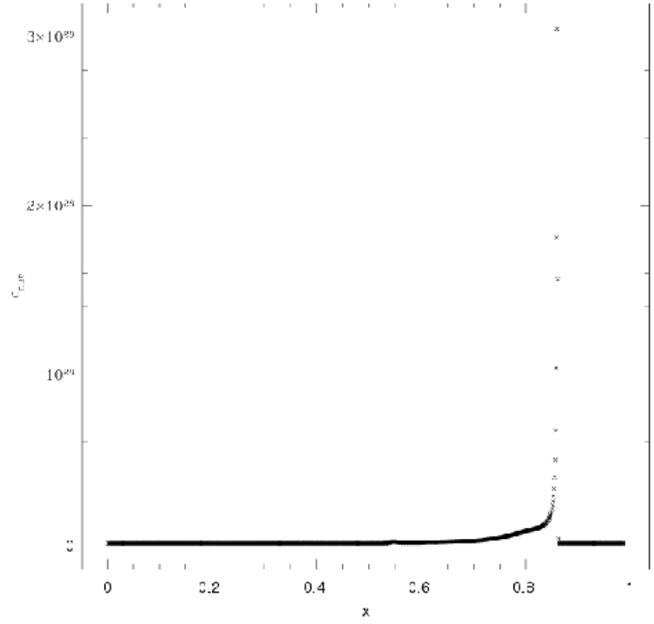}}
	\caption{Nuclear energy generation (erg$ \, $g$^{-1} \, $s$^{-1}$) profile of a detonation front in pure $ ^{12}$C at 
	$T=10^8 \, $K and $\rho= 10^8 \, $g$ \, $cm$^{-3}$ at time = $6 \, 10^{-10}$s 
	with a resolution of $10^{-3}$cm. $X$ is in cm.}
	\label{eps_nucl_C_1_screen}
	\end{figure}

  The profiles of temperature, velocity, density and pressure 
  at time t = $5  \, 10^{-6} \, $s for a resolution of 10 cm are presented 
  in Fig. \ref{therm_C_1e4_screen}. 
   The same profiles, and the nuclear energy generation
   profiles at time t = $6  \, 10^{-10} \, $s for a resolution of 0.1 cm are presented 
  in Figs. \ref{therm_C_1_screen} and \ref{eps_nucl_C_1_screen}. 
  Ninety percent of the nuclear energy is already 
  liberated over a distance of $10^{-2} \, $cm.
  
  If we want to study the propagation of the detonation around a neutron star surface, we will be mainly interested in the structure of 
  the end of the reaction zone. Indeed the detonation travels approximately $6 \, 10^6$ cm ($R_{NS} \approx 10 \, $km).  
  What happens on $10^{-2}$ cm is thus quite irrelevant, even if 90\% of the nuclear energy is alredy liberated there. 
  With a resolution of 
  only 10 cm, we can already infer the global properties of the detonation in spite of the fact that 
  we miss the destruction of C, the production and destruction of O, the production of Si and of S. All these 
  reactions occur within one computational cell just behind the shock, but this has no global impact on the mean thermodynamic 
  value at the end of the reaction length. This can be seen by comparing the thermodynamic profiles obtained with both resolutions (Figs. 
  \ref{therm_C_1e4_screen} and \ref{therm_C_1_screen}). However, fully resolving the detonation would be very important 
  for the study of the extinction of the 
  detonation (Maier \& Niemeyer \cite{Maier2006}). This question is not tackled here. 
  The profiles of Fig. \ref{therm_C_1e4_screen} 
  give 
  approximate values only, as the detonation is not perfectly resolved.

\begin{figure}
	\resizebox{\hsize}{!}{\includegraphics{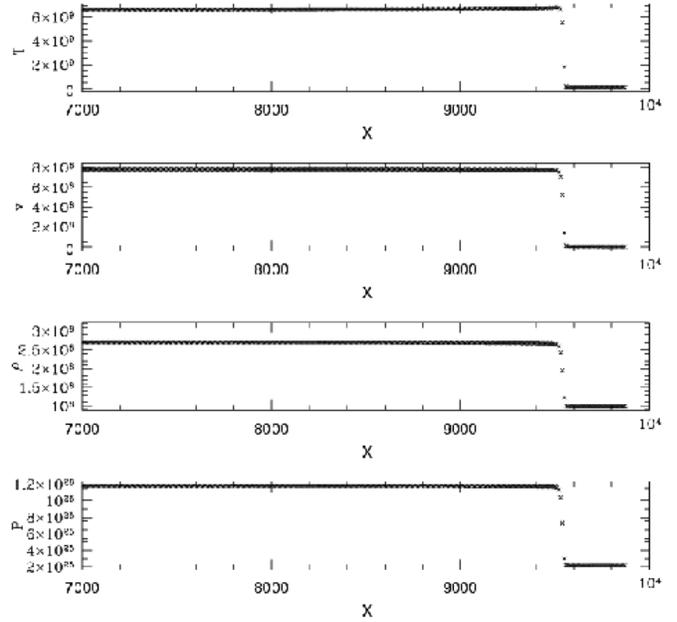}}
	\caption{Same as Fig. \ref{therm_C_1e4_screen}, but at time = $7 \, 10^{-6}$s and for a mixture 
	$X_{ ^{12}\mbox{C}} = 0.3$ and $X_{ ^{52}\mbox{Fe}} = 0.7$.}
	\label{therm_C_F_1e4_screen}
	\end{figure}

\begin{figure}
	\resizebox{\hsize}{!}{\includegraphics{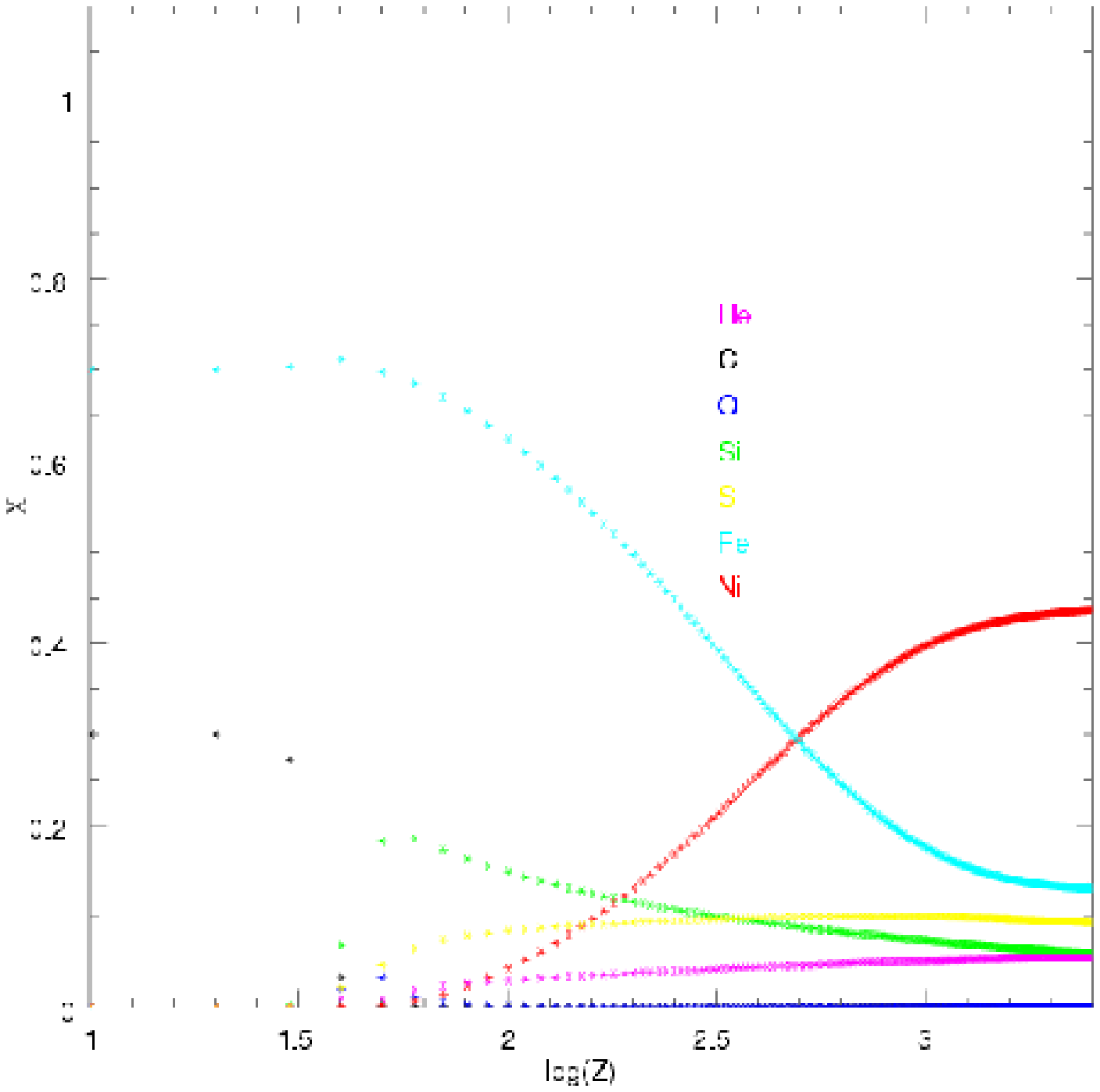}}
	\caption{Nuclear mass fraction profiles of a detonation front in a mixture $X_{ ^{12}\mbox{C}} = 0.3$
	and $X_{ ^{52}\mbox{Fe}} = 0.7$ at 
	$T=10^8 \, $K and $\rho= 10^8 \, $g$ \, $cm$^{-3}$ at time = $7 \, 10^{-6}$s. $Z$ is the distance to 
	the shock in cm.}
	\label{nucl_C_F_1e4_screen}
	\end{figure}

  A simulation with a composition $X_{ ^{12}\mbox{C}} = 0.3$ and $X_{ ^{52}\mbox{Fe}} = 0.7$ (SB02) 
  has also been conducted, the ignition conditions being 
  the same as for pure $ ^{12}\mbox{C}$. 
  The temperature, velocity, density and pressure 
  profiles at time t = $7  \, 10^{-6} \, $s are presented 
  in Fig. \ref{therm_C_F_1e4_screen}.
  In Fig. \ref{nucl_C_F_1e4_screen} 
  the nuclear mass fraction of some species is shown as a function of the distance to the shock. We have only performed 
  a simulation with a 10 cm resolution for the reasons explained above.
  
  From these two sets of results (detonation in pure $^{12}$C and in a mixture of $ ^{12}$C and $ ^{52}$Fe), we see that the composition of the 
  material before the passage of the detonation wave 
  affects the velocity of propagation of the detonation wave and the composition in 
  the burned material. For pure $ ^{12}$C the velocity of propagation of the detonation is $D \approx 1.3 \, 10^9 \, $cm s$^{-1}$, 
  and for the mixture of $ ^{12}$C and $ ^{52}$Fe, $D \approx 1.22 \, 10^9 \, $cm s$^{-1}$. The pure $ ^{12}$C detonation mainly produces 
  $ ^{4}$He, and the mixed detonation leads essentially to $ ^{56}$Ni.

%__________________________________________________________________
\section{Conclusions}\label{Conclusions}

   Our hydrodynamical algorithm for modeling astrophysical detonations has been shown to be robust to test cases. In particular 
   it reproduces quite well the steady state solution obtained with a totally different code. This give us confidence for future 
   multi-dimensional simulations. Some improvements still need to be made, like the inclusion of gravity, of non-ideal terms in the 
   equation of state, and more up-to-date data for the nuclear reaction network. 
   
    We have underlined the large difference 
   between the total reaction length and the length on which some species (e.g. carbon) burn in 
   conditions relevant to superbursts. This 
   difference leads to enormous numerical difficulties because all the length scales cannot be resolved at the same time during a 
   single simulation 
   (except possibly with some sub-grid models).
   We have shown that the carbon detonation in superburst conditions might be studied by a partial resolution approach.  
   This important, especially 
   in multidimensional simulations where the computation time is a crucial limitation.

   Our simulations give the global thermodynamic state of the material after the passage 
   of the detonation in the carbon layer at the surface of a neutron star. These 
   conditions are not very sensitive to the exact initial composition of the matter, in contrast to the final composition.
   In both cases, however, all the carbon is burned. It has to be replenished to allow the occurence 
   of a subsequent superburst.

   It is unlikely that the entire carbon layer ignites at the same time. More probably, ignition spots develop. 
   Since the detonation velocity depends on the composition of the carbon layer before ignition, the time for the 
   detonation to propagate laterally around the neutron star also depends on the composition.
   
   In a subsequent paper we will investigate a larger parameter space of thermodynamic condition. In particular, we
   will study the impact of the initial temperature or density. It would also be more realistic to introduce an initial temperature and density 
   profile. Two-dimentional superburst simulations are under study and will complement the work of Zingale et al. (\cite{Zingale2001}) 
   for normal type I X-ray bursts. 
   We will also investigate the vertical propagation of the detonation and its interaction with an overlying 
   helium layer. One of the aims of this simulation is to examine if the penetration of a detonation wave into the helium layer 
   could given rise to the precursor observed prior to the superburst from 4U 1820-30.
   
%__________________________________________________________________
\begin{acknowledgements}

      The numerical simulations were performed on the parallel computers of the Intensive Computing Storage 
      of UCL and on HYDRA, the new Scientific Computer Configuration at the VUB/ULB Computing Centre. 
      We are grateful to M. Arnould for a careful reading of the manuscript.
      We would like to thank J. Francois for the implementation of the stellar equation of state routines. 
      We are also grateful to the anonymous referee for his/her remarks and suggestion for future work.

\end{acknowledgements}
%__________________________________________________________________

\end{document}